\documentclass[symmetry,review,accept,pdftex,moreauthors]{Definitions/mdpi} 
\firstpage{1} 
\pubvolume{1}
\issuenum{1}
\articlenumber{0}
\pubyear{2022}
\copyrightyear{2022}
\externaleditor{Academic Editor: {Sergei D. Odintsov }} 
\datereceived{} 
\dateaccepted{} 
\datepublished{} 
\hreflink{https://doi.org/} 

\Title{Non-Thermal Fixed Points in Bose Gas Experiments}

\TitleCitation{Non-Thermal Fixed Points in Bose Gas Experiments}

\Author{Lucas 
 Madeira $^{1,}$*\orcidA{} and Vanderlei S. Bagnato $^{1,2}$}

\AuthorNames{Lucas Madeira  and Vanderlei S. Bagnato}

\AuthorCitation{Madeira, L.; Bagnato, V.S.}

\address{%
$^{1}$ \quad Instituto de F\'isica de S\~ao Carlos, Universidade de S\~ao Paulo, CP 369, 
S\~ao Carlos, 
 Brazil; vander@ifsc.usp.br\\
$^{2}$ \quad Hagler Fellow, 
 Hagler Institute for Advanced Study, Texas A\&M University, College Station, TX 77843, USA}

\corres{Correspondence: madeira@ifsc.usp.br}

\abstract{One of the most challenging tasks in physics has been understanding the route an out-of-equilibrium system takes to its thermalized state. This problem can be particularly overwhelming when one considers a many-body quantum system. However, several recent theoretical and experimental studies have indicated that some far-from-equilibrium systems display universal dynamics when close to a so-called non-thermal fixed point (NTFP), following a rescaling of both space and time. This opens up the possibility of a general framework for studying and categorizing out-of-equilibrium phenomena into well-defined universality classes. This paper reviews the recent advances in observing NTFPs in experiments involving Bose gases. We provide a brief introduction to the theory behind this universal scaling, focusing on experimental observations of NTFPs. We present the benefits of NTFP universality classes by analogy with renormalization group theory in equilibrium critical~phenomena.}


\begin{document}

\section{Introduction}


Although studying closed interacting quantum many-body systems is challenging, some theories have successfully described aspects of these systems~\cite{Polkovnikov2011}. Quantum statistical mechanics provides a good description of many physical systems in thermodynamic equilibrium. However, most natural phenomena occur under conditions outside equilibrium. This brings us to a new challenge in searching for descriptions of out-of-equilibrium conditions that allow a reasonable understanding of the~observations.

Some observables are insensitive to initial conditions and system parameters in certain~situations, leading to universal phenomena.
In the case of far-from-equilibrium initial conditions, this can be observed well before an equilibrium or a quasi-stationary state is reached. This has been investigated in a wide variety of systems such as
the inflation in the early Universe~\cite{Kofman1994,Micha2003,Berges2008,Orioli2015,Moore2016},
heavy-ion collisions producing quark--gluon matter~\cite{Baier2001,Berges2014,Berges2015,Berges2020},
and cold-gas systems,
in both theory~\cite{Orioli2015,Lamacraft2007,Barnett2011,Schole2012,Schmidt2012,Hofmann2014,Williamson2016,Chantesana2019}
and experiments~\cite{Erne2018,Prufer2018,Glidden2021,Garcia2021}.
A universal spatio-temporal scaling emerges, independent of the initial state or microscopic~parameters.

It has been suggested that the universal scaling observed in these far-from-equilibrium isolated systems is due to the presence of non-thermal fixed points (NTFPs)~\cite{Berges2008,Nowak2011,Orioli2015,Berges2015,Schole2012}.
In the literature, it is possible to find evidence of non-thermal {universality} classes encompassing a variety of systems. These are far away from any phase transitions; hence, their mechanism cannot be the same as that of the well-known critical phenomena and their characteristic exponents.
Another difference is that no fine-tuning of the parameters is required to observe the characteristic scaling provided by nearby NTFPs, unlike in equilibrium critical phenomena, where system variables need to be tuned to specific critical values.
This is evidenced by the universal scaling for several different initial states in NTFP~investigations.

Recently, with~the advent of Bose--Einstein condensates (BECs)~\cite{Pethick2008}, the~possibility of carrying out controlled experiments has become a reality, opening a window of opportunity to investigate intrinsic properties of these systems or even classify them according to their pattern of time~evolution.

The production of an out-of-equilibrium BEC can be achieved experimentally in several ways, one of them being via so-called ``quantum quenching''. The~system, initially in equilibrium, is represented by a Hamiltonian $H(\alpha_0)$, where $\alpha_0$ is a set of system parameters. Then, an~abrupt change is made to a new situation described by a Hamiltonian $H(\alpha_1)$, where~$\alpha_1$ is a new collection of parameters. In~this case, the~system leaves the equilibrium condition, having its previous state determined by $H(\alpha_0)$ as its initial condition. Its temporal evolution is unitary and governed by the new Hamiltonian $H(\alpha_1)$. The~operator $U= \exp\left[-i H(\alpha_1)t/\hbar \right]$ determines what happens until the system can decay to a new state of $H(\alpha_1)$. One example corresponds to abrupt potential changes, when only a few degrees of freedom are present, allowing a simplified system evolution. This simplicity can change when dealing with a system with many degrees of freedom, which is the case in a many-body quantum system. Each subsystem can see the rest as a reservoir, and~the final time evolution can take quite complicated routes. There are many open questions, including the fundamental question of how the system evolves temporally toward equilibrium, or~even how to quantify the out-of-equilibrium state and~identify which system components are determinant in the typical relaxation time for a new state of~equilibrium.

{Since there are many quench protocols that can be employed to produce out-of-equilibrium BECs, some even preserving symmetries present in the equilibrium state, the~question of whether a system can approach a NTFP arises. In~general, this depends strongly on the chosen initial conditions, which should correspond to extreme out-of-equilibrium configurations}~\cite{Schmied2019}.
{For example, in~the case of a dilute Bose gas, this could be achieved by populating only modes below a certain momentum scale $Q$. The~initial state, a~constant momentum distribution for $k<Q$ that drops abruptly at $k=Q$, is strongly overpopulated at low momenta (compared to the final equilibrium distribution). Hence, the~subsequent dynamics is of particle transport toward lower momenta and energy migration to high momenta. If~the spatio-temporal scaling presented in Section~}\ref{sec:universal}{ is observed, then this is a ``smoking gun'' indicating a nearby NTFP}~\cite{Schmied2019}.

In experiments, preparing and maintaining a closed system can be very challenging. Cold-atom systems are, in~this sense, very close to the ideal. Isolation allows experimental access by external agents in a controlled manner. The~isolation of trapped condensates and the excellent control of quantum states are fundamental elements required to achieve long coherence times, an~essential feature in out-of-equilibrium studies since the coherence time must be of at least the same order as the temporal scale of the dynamics~involved.

Some clarification about the scope of this review is in order.
We chose to focus our attention on experiments involving Bose gases~\cite{Erne2018,Prufer2018,Glidden2021,Garcia2021}.
Although this corresponds only to a small part of all research being conducted on NTFPs, we consider these examples to be illustrative of the field. Notably, it is possible to present straightforward applications of the theory without exploring its inner workings. For~readers that wish to do so, we provide references that may be helpful~\cite{Nowak2013,Chantesana2019,Schmied2019}.
One of the motivations is to use cold-atom setups to simulate the dynamics of currently inaccessible systems; for example, the inflation of the early Universe. This is counter-intuitive because cold-atom systems are in the low-energy regime, whereas the early Universe is at the other end of the energy spectrum. It is only possible through the universal scaling due to the presence of~NTFPs.

This paper is organized as follows. In~Section~\ref{sec:universal}, we introduce the universal scaling provided by NTFPs and related quantities of interest.
Section~\ref{sec:experiments} presents four experiments dealing with NTFPs in Bose gas systems:
a quasi-one-dimensional Bose gas in Section~\ref{sec:1d};
a spin-1 condensate in Section~\ref{sec:spinor};
a homogeneous condensate in three-dimensions in Section~\ref{sec:homogeneous};
and a turbulent harmonically trapped BEC in Section~\ref{sec:turbulence}.
Finally, a~summary is given in Section~\ref{sec:final}.


\section{Universal Scaling and~NTFPs}
\label{sec:universal}

Our goal in this section is to present the universal scaling function and its associated exponents. The~motivations, details, and~derivations regarding the NTFP theory can be found in~\cite{Nowak2013,Chantesana2019,Schmied2019} and the references therein.
The theory developed for NTFPs is inspired by equilibrium renormalization group theory where, in~the vicinity of a phase transition, the~correlations are self-similar (independent of the resolution). Thus, scaling the spatial resolution by a parameter $s$ creates a correlation function that depends on the distance $x$ between two points behave according to $C(x;s)=s^\zeta f(x/s)$. In~this way, the~correlation is characterized exclusively by a universal exponent $\zeta$ and a function $f$.
A fixed point in equilibrium critical phenomena corresponds to a situation where varying $s$ does not change $C(x;s)$, meaning that the universal function is a power law, $f(x) \propto x^{-\zeta}$. In~physical systems, this behavior is approximate, and information may be retained about characteristic~scales.

\begin{figure}[H]

\includegraphics[width=0.45\linewidth]{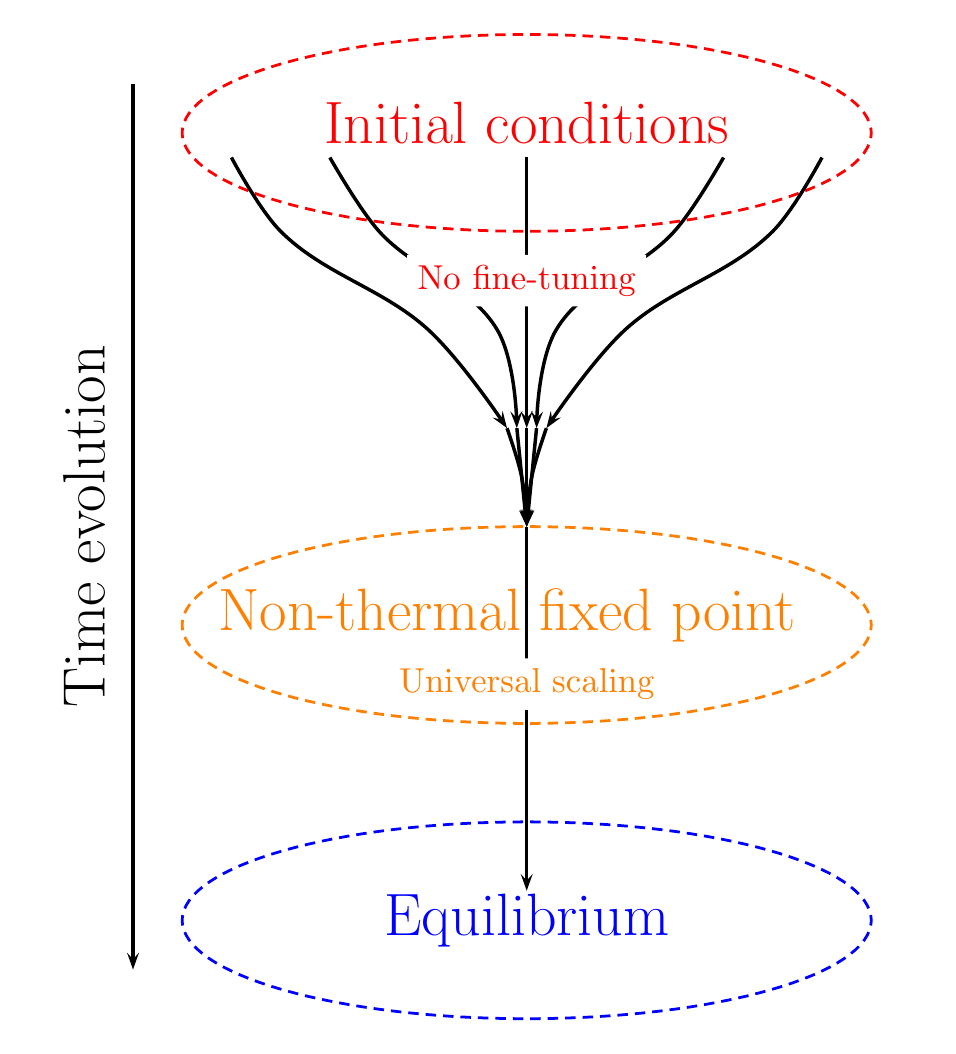}
\caption{
Illustration of the dynamics of a system passing near an NTFP.
For several initial conditions (the key idea being that no fine-tuning is needed), the~system can pass near a non-thermal fixed point.
The correlation functions show a spatio-temporal scaling with a universal function when that occurs. After~some time, the~system leaves the vicinity of the NTFP and reaches equilibrium.
}
\label{fig:universal}
\end{figure}

Non-thermal fixed points in the dynamics of far-from-equilibrium systems are analogous to fixed points in critical equilibrium phenomena, with~the distinction that the former uses the time $t$ as the scale parameter. Near~NTFPs, the~correlations take the form $C(x,t)=t^\alpha f(t^{-\beta}x)$, now with two universal exponents.
Universality classes, in~this context, correspond to the same exponents $\alpha$ and $\beta$ and a scaling function $f$.
Figure~\ref{fig:universal} shows a schematic time evolution for a system that encounters an~NTFP.

In the case of cold bosonic gases, the~momentum distribution can be used to characterize the time evolution of the system.
Consider an isotropic momentum distribution, i.e.,~a distribution that is only a function of
$k=|\textbf{k}|$ and time, $n(k,t)$. The~total number of particles $N(t)$ is obtained by integrating the momentum distribution,
\begin{equation}
\label{eq:norm}
N(t)=\int d^dk \ n(k,t).
\end{equation}

We provide the $d$-dimensional expression, since we cover the cases $d=1,2,$ and 3 in this review.
The fluctuations in the total number of particles in cold-gas experiments are relatively small due to the low temperatures involved and the high levels of control that one can exert over these systems.
However, unwanted losses occur mainly due to the heating of the sample.
For this reason, several experimental studies report the momentum distribution normalized by the total number of particles at a given time, $\bar{n}(k,t)\equiv n(k,t)/N(t)$.
In theoretical and numerical investigations, $N$ is a fixed number and~this distinction is~irrelevant.

The universal scaling displayed by a system in the vicinity of an NTFP is given by
\begin{equation}
\label{eq:scaling}
\bar{n}(k,t)=\left(\frac{t}{t_0}\right)^\alpha f\left[\left(\frac{t}{t_0}\right)^\beta k\right].
\end{equation}
where
$\alpha$ and $\beta$ are scaling exponents, $f$ is a universal scaling function, and~$t_0$ is an arbitrary reference time within the time interval where the scaling takes place.
From this definition of the scaling exponents, it is possible to see that the sign of $\alpha$ is related to the direction of the particle transport. Positive values indicate particles migrating toward lower momenta, whereas negative values increase the population of the high-momentum~components.

\subsection{Global~Observables}
\label{sec:global}

Besides the important result of Equation~(\ref{eq:scaling}), there are also global observables of interest when dealing with systems that display universal scaling due to the presence of nearby NTFPs.
The main difference is that these quantities are computed in the infrared (IR) scaling region, $k_D\leqslant (t/t_0)^{\beta} k\leqslant  k_c$, where $k_D$ is the smallest wave vector that can be measured (inversely proportional to the largest length scale of the system $D$, typically its size) and~$k_c$ defines the cut-off where the universal scaling takes~place.

One of the observables is the number of particles in the scaling region,
\begin{equation}
\label{eq:N}
\bar{N}(t)=\int\limits_{k\leqslant\left(\frac{t}{t_0}\right)^{-\beta}k_c} d^dk \ \bar{n}(k,t) \propto \left(\frac{t}{t_0}\right)^{\alpha- d \beta}.
\end{equation}

A necessary condition for Equation~(\ref{eq:scaling}) to hold is that $\bar{N}(t)$ is constant during the time interval where the universal scaling is observed.
The time dependence $\bar{N}(t)\propto t^{\alpha-d\beta}$ can be derived straightforwardly by inserting Equation~(\ref{eq:scaling}) into Equation~(\ref{eq:N}) and changing the variables of~integration.

The moments of the momentum distribution can also be computed in the region of the universal scaling. The~second moment has the physical interpretation of the mean kinetic energy per particle in the scaling region,
\begin{equation}
\label{eq:M2}
\bar{M}_2(t)=\int\limits_{k\leqslant\left(\frac{t}{t_0}\right)^{-\beta}k_c}d^dk \ \frac{k^2 \bar{n}(k,t)}{\bar{N}(t)} \propto \left(\frac{t}{t_0}\right)^{-2\beta}.
\end{equation}

Its time dependence, $\bar{M}_2\propto t^{-2\beta}$, can also be derived using Equations~(\ref{eq:scaling}) and (\ref{eq:N}). Hence, the~sign of $\beta$ is related to the direction of the energy transport. For~$\beta>0$, the~energy leaves the scaling region and migrates to higher momenta, whereas a negative sign indicates an energy increase in the IR scaling~interval.

These are not the only global observables that can be defined.
\textls[-15]{For example, in~\mbox{Section~\ref{sec:spinor}} we encounter another quantity, relevant in the context of a spinor~condensate.}

\section{Experiments}
\label{sec:experiments}
\unskip

\subsection{One-Dimensional Bose~gas}
\label{sec:1d}

The authors in~\cite{Erne2018} produced a one-dimensional Bose gas by strongly quenching a three-dimensional one.
They reported a universal scaling in the time-dependent momentum distribution due to the presence of an~NTFP.

These authors employed $^{87}$ Rb atoms, which correspond to a repulsively interacting Bose gas in~a very elongated (quasi-1D) harmonic trap.
At the beginning of the experiment, the~thermal gas was just above the critical temperature.
During the quench, the~trap depth was ramped linearly, such that its final value was below the first radially excited state. This causes the evaporation of atoms occupying higher energy states. Finally, the~depth of the trap was raised to close the trap. The~resulting Bose gas was in a far-from-equilibrium condition, and~its evolution was recorded after a time $t$.
They were able to probe the system using both the {in situ} density and the momentum distribution, which is the quantity of interest for the universal scaling in the form of Equation~(\ref{eq:scaling}).

Some aspects of the momentum distribution at early and late times were known.
The authors provided a quantitative description of the initial state in terms of solitonic defects~\mbox{\cite{Schmidt2012,Karl2017}}. Since the quenching procedure is almost instantaneous, the~initial state had a large population of high-energy modes, which makes the observation of the universal dynamics associated with NTFPs possible when the system relaxes.
During the time evolution of the system, a~peak appears at low momenta, indicative of the formation of the quasi-condensate. 
The system reaches a thermal quasi-condensate state at late times, which is described by a Lorentzian function (its width is inversely proportional to the temperature).

\begin{figure}[H]
\includegraphics[width=\linewidth]{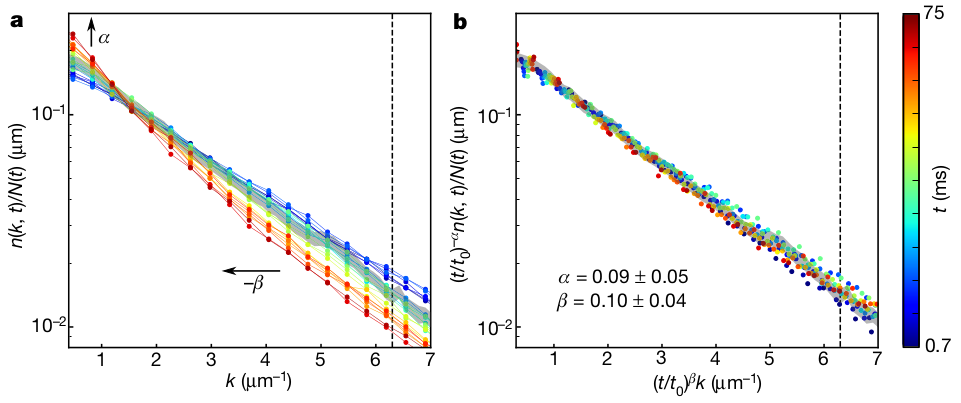}
\caption{Universal 
 scaling dynamics observed by the authors of~\cite{Erne2018}. (\textbf{a}) Time evolution of the normalized momentum distributions.
(\textbf{b}) Momentum distributions scaled according to Equation~(\ref{eq:scaling}). All the curves collapse into a single function, signaling the universal scaling.
Reprinted by permission from Springer Nature Customer Service Centre GmbH: Springer Nature, Nature, Erne et al., Universal dynamics in an isolated
one-dimensional Bose gas far from
equilibrium, \copyright \ 2018.
}
\label{fig:erne}
\end{figure}

The normalized momentum distributions are shown in Figure~\ref{fig:erne}a. As~time progresses, the~distribution increases (decreases) at low (high) momenta, signaling the formation of the quasi-condensate.
In Figure~\ref{fig:erne}b, the~authors provide the scaled curves using Equation~(\ref{eq:scaling}) with $\alpha=0.09(5)$ and $\beta=0.10(4)$. All curves, below~a cut-off value $k_c$ indicated by a dashed line in the figure, collapse into a single universal function.
Both exponents are positive, indicating that particles migrate toward the low-momenta region, and~energy flows in the high-momenta direction. This is consistent with the formation of a quasi-condensate after the quenching procedure is performed.
The theoretical study in~\cite{Orioli2015} predicts a value of $\beta=1/2$, independent of the dimension $d$, for~far-from-equilibrium dynamics in an isolated Bose gas following a strong quench.
The authors of~\cite{Erne2018} provide arguments for why this theory does not apply fully to their~system.
As well as analyzing the momentum distributions, the~authors also computed the number of particles and the mean kinetic energy per particle in the scaling region (Equations~(\ref{eq:N}) and (\ref{eq:M2})).
The quantity $\bar{N}$ was approximately constant in that region, which is consistent with the time-dependence prediction of Equation~(\ref{eq:N}), i.e., $\bar{N}(t)\propto t^{\alpha-d\beta}$, since $d=1$ and the authors found $\alpha\approx\beta$.
Moreover, the~$\bar{M}_2\propto t^{-2\beta}$ behavior was also verified in the region of interest.
As expected, both quantities showed a different time dependence outside the scaling~region.

\subsection{Spinor Bose~Gas}
\label{sec:spinor}

The authors in~\cite{Prufer2018} observed universal dynamics in a quasi-one-dimensional spinor Bose gas~\cite{Sadler2006} by analyzing spin correlations.
They employed a $^{87}$Rb gas in the $F=1$ hyperfine state, which has three possible magnetic sublevels: $m_F=-1,0,1$. Hence, it behaves as a spin-1 system with ferromagnetic interactions~\cite{Kurn2013}.
Initially, all atoms are in the $m_F=0$ state.
The system is driven far from equilibrium by a sudden change in the energy splitting of the sublevels, producing excitations in the $F_x-F_y$ plane.

Although both this system and the one presented in Section~\ref{sec:1d} are quasi-one-dimensional systems, the~degrees of freedom and underlying physics are quite different. Hence, the~function entering into the scaling described by Equation~(\ref{eq:scaling}) is not as straightforward as being simply the momentum distribution.
For details regarding the function and the measurements, the~reader is referred to \cite{Prufer2018}.
Here, we present only a brief~overview.

First, the authors computed the mean spin length,
$\langle |F_\perp (t)| \rangle$, where $F_\perp=F_x+iF_y$.
A local angle was defined and extracted using
$\theta(y,t)=\arcsin(F_x(y,t)/\langle |F_\perp (t)| \rangle)$.
Fluctuations were probed by means of a two-point correlation function, $C(y,y';t)=\langle\theta(y,t)\theta(y',t)\rangle$.
Finally, the~desired function is the structure factor, which is the averaged Fourier transform:
\begin{equation}
\label{eq:sf}
f_\theta(k,t)=\int dy \int d\bar{y} \ C(y+\bar{y}
,y;t)e^{-2\pi i k\bar{y}}.
\end{equation}

Although the derivation of the function defined in Equation~(\ref{eq:sf}) may seem more intricate than the more familiar momentum distribution of Section~\ref{sec:1d},
both are functions of momentum and time and are determined by experimental parameters and the initial~state.

\begin{figure}[H]
\includegraphics[width=\linewidth]{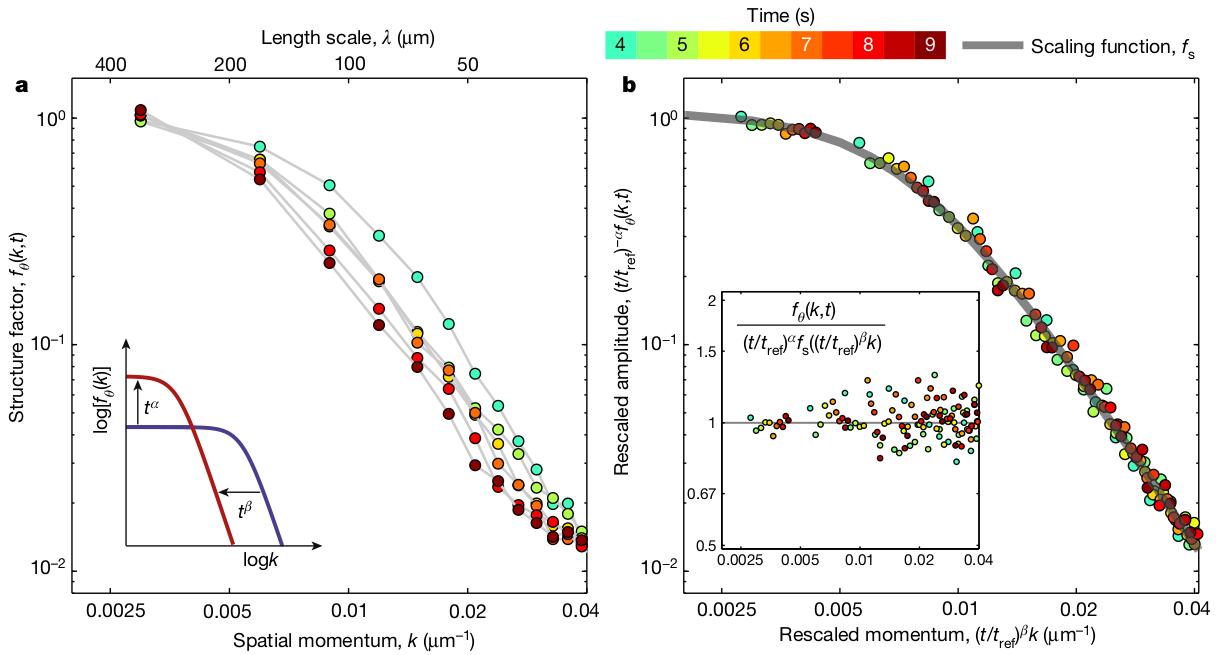}
\caption{Universal 
 scaling in a spinor Bose gas~\cite{Prufer2018}. (\textbf{a}) Time evolution of the structure factor (Equation~(\ref{eq:sf})).
(\textbf{b}) After the scaling of Equation~(\ref{eq:scaling}) has been applied, the~curves collapse into a universal function.
Reprinted by permission from Springer Nature Customer Service Centre GmbH: Springer Nature, Nature, Pr\"ufer et al., Observation of universal dynamics in
a spinor Bose gas far from
equilibrium, \copyright \ 2018.
}
\label{fig:prufer}
\end{figure}

In Figure~\ref{fig:prufer}a the authors present the time evolution of the structure factor. It is possible to see a shift toward lower momenta as time passes.
When the scaling of Equation~(\ref{eq:scaling}) is applied to the data, all the points collapse into a single universal curve, as shown in Figure~\ref{fig:prufer}b.
The exponents employed were $\alpha=0.33(8)$ and $\beta=0.54(6)$.
The authors also find a relevant global observable $\int dk f_\theta(k,t)$, which is an~approximately conserved quantity during the evolution.
The arguments used in Section~\ref{sec:global} can be employed to derive its time dependence $\propto t^{\alpha-\beta}$.
The observed migration of the conserved quantity toward lower momenta is consistent with $\beta>0$.

Although the values of the exponents do not strictly correspond to $\alpha\approx \beta$, as~one would expect from $\alpha-d\beta=0$, we should keep in mind that this system is composed of $N_s=3$ unidimensional Bose gases, for~which there are no theoretical predictions.
For other systems described by $\mathcal{O}(N_s)$ symmetric models and $d\geqslant 2$, the~universal value of $\beta\approx 0.5$ has been predicted~\cite{Orioli2015}.


\subsection{Homogeneous Three-Dimensional Bose~Gas}
\label{sec:homogeneous}

The authors of~\cite{Glidden2021} observed a bidirectional dynamical scaling in a homogeneous three-dimensional Bose gas~\cite{Gaunt2013}.
While the systems described in Sections~\ref{sec:1d} and \ref{sec:spinor} only report universal scaling in the IR region, the~authors of~\cite{Glidden2021} also observed scaling in the ultraviolet (UV) region of the momentum distribution, hence the term bidirectional.
{Although the authors found a different set of exponents for the universal scaling in each region, these do not correspond to two different NTFPs. Instead, they are a consequence of particle transport toward small momenta and energy transport toward large momenta, in~agreement with NTFP theory} \cite{Schmied2019}.

In their experiment, the authors employed $^{39}$K atoms in a cylindrical box, producing a homogeneous 3D Bose gas.
Their experimental protocol depended on the depth of the optical box trap and the $s$-wave two-body scattering length $a$, which characterizes the interparticle interactions.
First, a~cloud containing the atoms just above the condensation temperature was prepared.
The idea was to quickly remove atoms and energy from the system to produce a far-from-equilibrium state.
This was achieved by turning off the interactions ($a\to 0$) and lowering the depth of the trap so that high-energy atoms evaporated. Since there are no interactions, the~system does not thermalize.
This step removed $\approx$77\% of the atoms and $\approx$98\% of the energy such that, if~it were in equilibrium, a~significant fraction of the system would condense.
Next, the~system was closed again by increasing the depth of the trap potential.
At $t=0$, interactions were turned on again ($a\ne 0$) and thermalization began to take place.
After a variable time $t$, the~momentum distribution $n(k,t)$ was obtained through absorption images.
For long enough times, the~system reaches equilibrium with both condensed and thermal~components.

During thermalization, the~total number of particles $N$ and total energy $E$ are conserved, but~there are two distinctive flows in opposite directions as time progresses. The~majority of particles migrate toward the IR region, consistent with the condensate formation, but~a small fraction of the atoms transfer the energy in the UV~direction.

\begin{figure}[H]
\includegraphics[width=\linewidth]{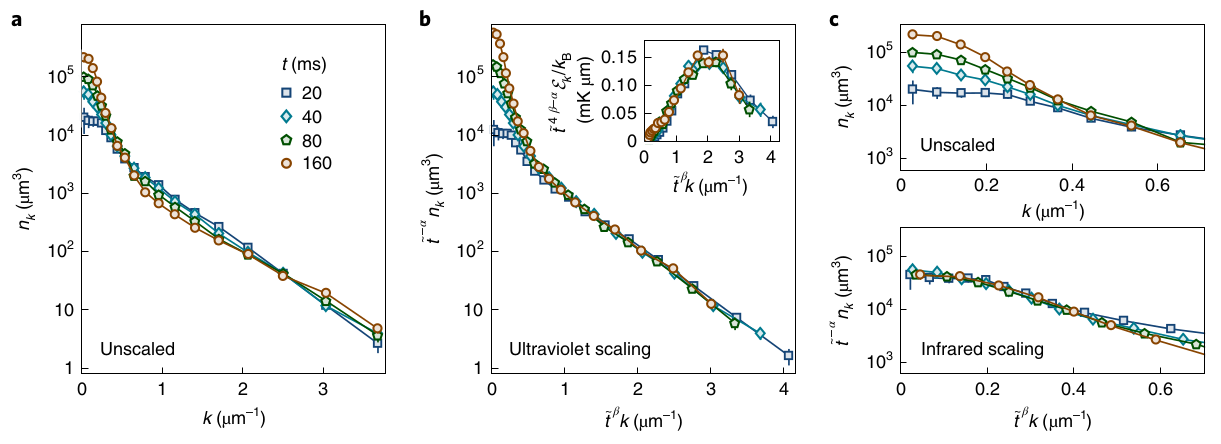}
\caption{Universal 
 bidirectional scaling in a homogeneous Bose gas~\cite{Glidden2021}. (\textbf{a}) Momentum distributions as a function of time.
(\textbf{b}) Scaling provided by Equation~(\ref{eq:scaling}) with $\alpha=-0.70(7)$ and $\beta=-0.14(2)$, which collapses the curves into a universal function for the UV region.
(\textbf{c}) The top panel shows the low-momenta region of the momentum distributions, while the bottom one depicts the scaling of Equation~(\ref{eq:scaling}) with $\alpha=1.15(8)$ and $\beta=0.34(5)$.
Reprinted by permission from Springer Nature Customer Service Centre GmbH: Springer Nature, Nature Physics, Glidden et al., Bidirectional dynamic scaling in an
isolated Bose gas far from
equilibrium, \copyright \ 2021.
}
\label{fig:glidden}
\end{figure}

The presence of a nearby NTFP allows for the universal scaling of Equation~(\ref{eq:scaling}) in a time interval between the initial and equilibrium states.
In Figure~\ref{fig:glidden}a, the~authors show the unscaled $n(k,t)$ profiles.
The UV region is described well by the exponents
$\alpha=-0.70(7)$ and $\beta=-0.14(2)$ (Figure~\ref{fig:glidden}b), while the IR region collapses into a universal function with 
$\alpha=1.15(8)$ and $\beta=0.34(5)$
(Figure~\ref{fig:glidden}c).

For a $d$-dimensional system with a dispersion relation of the form $\omega(k) \propto k^z$, energy density conservation requires $\alpha/\beta=d+z$ 
\cite{Chantesana2019,Schmied2019}.
Hence, the~UV scaling, where $\alpha/\beta \approx 5$, is consistent with energy-conserving transport with a quadratic dispersion relation in 3D.
Moreover, weak-wave turbulence predicts a value of $\beta=-1/6$ for the UV dynamics~\cite{Zakharov1992,Dyachenko1992}, close to that observed in~\cite{Glidden2021}.

For the IR region, $\alpha/\beta\approx 3$, consistent with particle conservation in $d=3$ dimensions.
While several theoretical investigations predict $\beta=1/2$ for the IR region~\cite{Berges2008,Orioli2015,Chantesana2019}, certain conditions may yield $\beta=1/3$~\cite{Schmied2019,Mikheev2019}.

The authors of~\cite{Glidden2021} also investigated a dynamical scaling depending on the interactions $t\to t a/a_0$, where $a_0$ is a reference scattering length.
The exponents obtained by this other scaling were similar to those previously obtained.

\subsection{Harmonically Trapped Three-Dimensional Bose~gas}
\label{sec:turbulence}

The authors in~\cite{Garcia2021} investigated the emergence of universal scaling due to the presence of NTFPs in~a harmonically trapped three-dimensional Bose gas, driven to a turbulent~state.

The experiment began with a cigar-shaped $^{87}$Rb BEC with a condensed fraction of $\approx$70\% in equilibrium.
The production of a far-from-equilibrium state was achieved by a sinusoidal time-varying magnetic field gradient, such that it was not aligned with the principal axes of the trap. This corresponds to rotations and distortions of the original trap shape.
The amplitude, frequency, and~duration of the excitation could be varied and controlled. At~$t=0$ the excitation was turned off, and~the system was left to evolve in the trap. After~a time $t$, absorption images were taken to obtain the momentum distribution of the~system.

\begin{figure}[H]
\includegraphics[width=\linewidth]{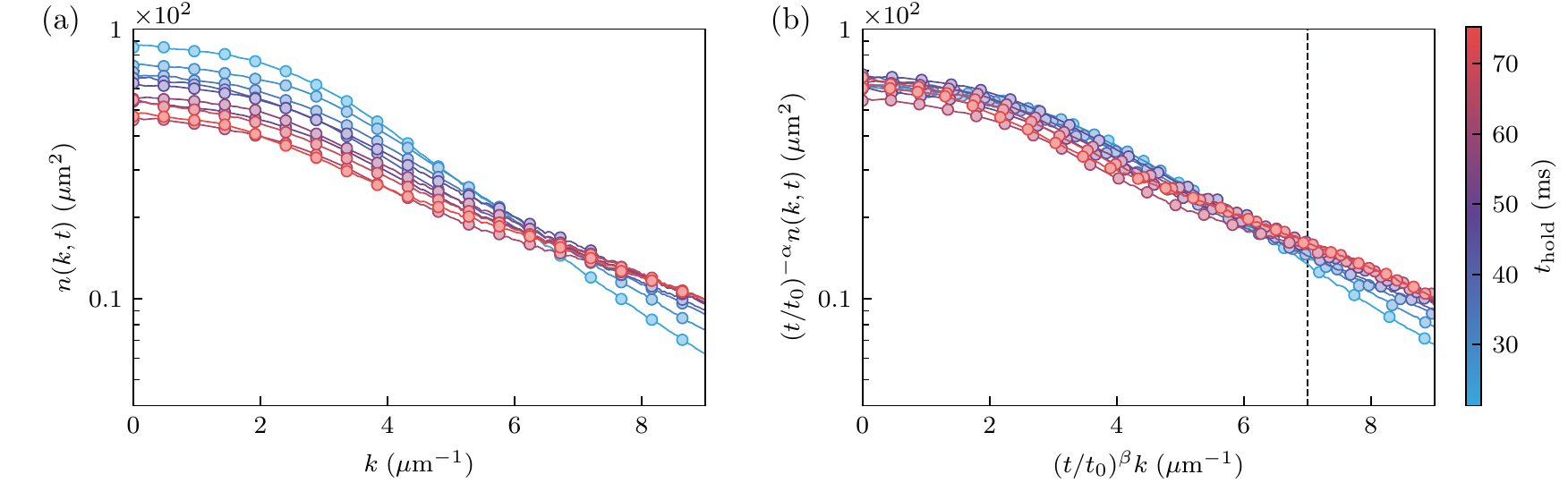}
\caption{Universal 
 scaling in a turbulent harmonically trapped Bose gas~\cite{Garcia2021}. (\textbf{a}) Momentum distributions of the turbulent state. (\textbf{b}) Scaled momentum distributions with $\alpha=-0.50(8)$ and $\beta=-0.2(4)$. Figure taken from~\cite{Garcia2021}.
}
\label{fig:garcia2d}
\end{figure}

The emergence of a turbulent state depends on the parameters of the excitation protocol, as identified by~a characteristic power-law behavior. Figure~\ref{fig:garcia2d}a shows the time evolution of the momentum distribution of a turbulent state. As~time passes, the~distribution shifts toward high momenta, indicating the depletion of the condensate. The~scaling employing Equation~(\ref{eq:scaling}) with $\alpha=-0.50(8)$ and $\beta=-0.2(4)$ is shown in Figure~\ref{fig:garcia2d}b, which collapses all curves into a single~function. 

\begin{figure}[H]
\includegraphics[width=\linewidth]{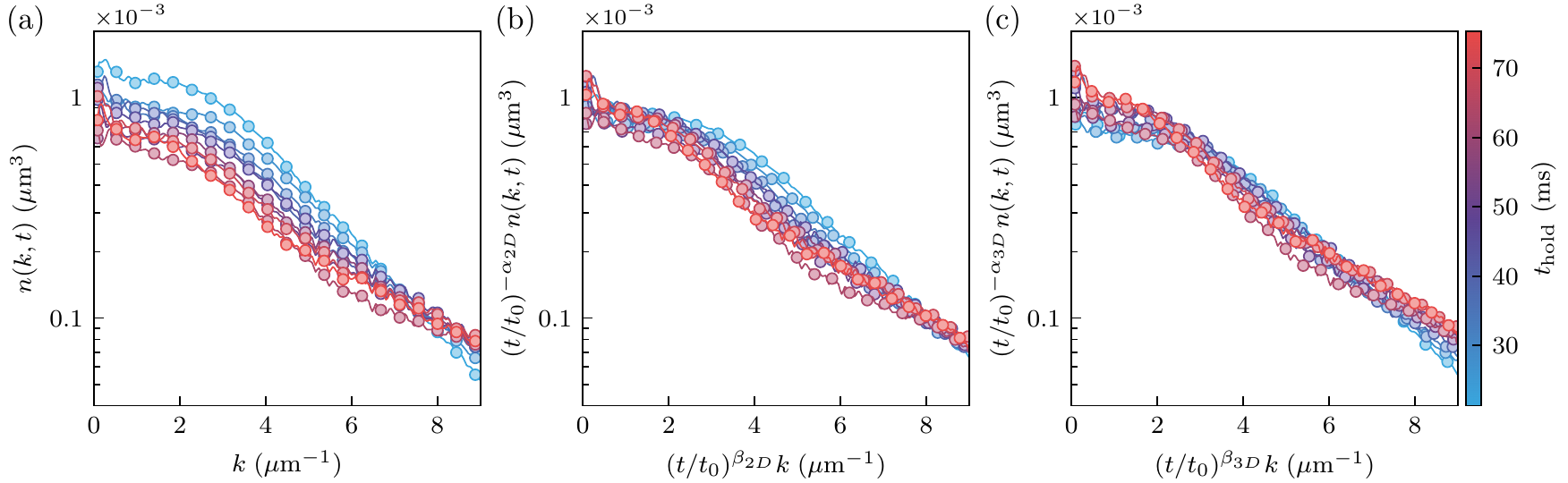}
\caption{(\textbf{a}) Three-dimensional 
 momentum distribution reconstructed with the inverse Abel transform. (\textbf{b}) Scaling provided by Equation~(\ref{eq:scaling}) with $\alpha=-0.50$ and $\beta=-0.2$, which are the~same exponents as those employed in the two-dimensional projection. (\textbf{c}) Scaling using $\alpha=-0.75$ and $\beta=-0.2$, corresponding to the prediction of the exponents for the three-dimensional case. The~collapse is much better than that shown in panel (\textbf{b}).
The figure is taken from~\cite{Garcia2021}.
}
\label{fig:garcia3d}
\end{figure}

The momentum distributions of Figure~\ref{fig:garcia2d}a were obtained via absorption images of the cloud, corresponding to two-dimensional projections of a three-dimensional system.
Using the inverse Abel transform~\cite{Bracewell1986} and some assumptions, the~authors in~\cite{Garcia2021} were able to reconstruct the three-dimensional momentum distributions (see Figure~\ref{fig:garcia3d}a).
The exponents determined using the two-dimensional projections did not provide the universal scaling, as~evidenced by Figure~\ref{fig:garcia3d}b.
The authors showed that
the exponents obtained through the scaling of a projection were related to those of the isotropic three-dimensional distribution through $\alpha_{\rm 3D}=3\alpha/2$ ($\beta$ remains the same). Using $\alpha=-0.75$ and $\beta=-0.2$, the~collapse into a universal function was much better adjusted, as~shown in Figure~\ref{fig:garcia3d}c.

The global observables of Equations~(\ref{eq:N}) and (\ref{eq:M2}) were also computed in~\cite{Garcia2021}. The authors observed a slight decrease in the particle number in the scaling region, $\bar{N}\propto t^{-0.1}$, and~an increase in the mean kinetic energy consistent with $\bar{M}_2\propto t^{-2\beta}$.
These observations are consistent with the energy leaving the IR region and the depletion of the condensate.
The authors also presented the benefits of merging quantum turbulence phenomena into a universality class of dynamically scaling systems characteristic of NTFPs~\cite{Orioli2015,Schole2012,Chantesana2019,Nowak2011,Karl2017,Schmied2019,Dyachenko1992,Nowak2012,Nowak2013,Scheppach2010}.

\section{Final~Remarks}
\label{sec:final}

In this review, we focused on experiments with Bose gases that, due to the presence of NTFPs, display the spatio-temporal scaling of Equation~(\ref{eq:scaling}). Table~\ref{tab:exponents} contains a summary of the systems and related exponents.
Although we discussed these experiments only in~terms of dimensionality, the~number of components, and the~signs of the exponents, these systems cover a wide range of~scenarios.

\begin{table}[H]
\caption{Summary 
 of the exponents $\alpha$ and $\beta$ found in the experiments covered in this review~\cite{Erne2018,Prufer2018,Glidden2021,Garcia2021}.
We indicate the dimensionality $d$, the~number of components $N_s$, and~the two different sets of exponents found in~\cite{Glidden2021,Garcia2021}.
}
\label{tab:exponents}

\newcolumntype{C}{>{\centering\arraybackslash}X}
\begin{tabularx}{\textwidth}{m{5cm}<{\centering}m{2.5cm}<{\centering}m{0.7cm}<{\centering}m{0.7cm}<{\centering}m{1.3cm}<{\centering}m{1.2cm}<{\centering}}
\toprule
\textbf{Bose Gas} &   & \boldmath{$d$} & \boldmath{$N_s$} & \boldmath{$\alpha$} & \boldmath{$\beta$} \\ \midrule
1D~\cite{Erne2018}              &    & 1 & 1 & 0.09(5) & 0.10(4) \\ \midrule
Spinor~\cite{Prufer2018}        &    & 1 & 3 &  0.33(8) & 0.54(6) \\ \midrule
\multirow{2}{*}{\vspace{-12pt}3D Homogeneous~\cite{Glidden2021}}                  & IR region & 3 & 1 & 1.15(8) & 0.34(5) \\
                                & UV region & 3 & 1 &  $-$0.70(7) & $-$0.14(2) \\ \midrule
\multirow{2}{*}{Turbulent, harmonically trapped~\cite{Garcia2021}} & 2D projection & 2 & 1 & $-$0.50(8)  & $-$0.2(4) \\
                                & 3D reconstruction & 3 & 1 & $-$0.75 & $-$0.2 \\\bottomrule

\end{tabularx}
\end{table}

A common aspect of the experiments presented in this review is the relation followed by the scaling exponents, $\alpha\approx d\beta$, in~the IR region.
This is not surprising, since according to Equation~(\ref{eq:N}), $\bar{N}\propto t^{\alpha-d\beta}$ and $\bar{N}(t)$ must be conserved during the interval where the scaling occurs.
Nevertheless, verifying these theoretical predictions experimentally in systems that are so different from each other strengthens the claim regarding~NTFPs.

In equilibrium critical phenomena, renormalization group theory and fixed points lead to critical exponents~\cite{Wilson1975}, which provide a unified description in terms of universality classes sharing the same exponents.
The experimental evidence provided by the studies presented in this review is crucial to providing something similar in isolated far-from-equilibrium systems.
Hopefully, the~interplay between theory and experiments may lead to classification schemes based on universal properties, which would be important for various systems.
This would allow, for~example, cold-atom experiments to be employed to simulate different systems of the same universality~class.

\vspace{6pt} 



\authorcontributions{
Conceptualization, V.S.B.;
methodology, L.M.;
software, L.M.;
validation, L.M.;
formal analysis, L.M.;
investigation, L.M.;
resources, V.S.B.;
data~curation, L.M.;
writing---original draft preparation, L.M.;
writing---review and editing, V.S.B.;
visualization, L.M.;
supervision, V.S.B.;
project administration, L.M.;
funding acquisition, V.S.B.
All authors have read and agreed to the published version of the~manuscript.}

\funding{
This work was supported by
the S\~ao Paulo Research Foundation (FAPESP)
under the grants 2013/07276-1, 2014/50857-8, and~2018/09191-7, and~by the
National Council for Scientific and Technological Development (CNPq)
under the grant 465360/2014-9.}

\institutionalreview{Not applicable.}

\informedconsent{Not applicable.}

\dataavailability{No new data were created or analyzed in this study. Data sharing is not applicable to this article.}


\conflictsofinterest{The authors declare no conflict of~interest.} 


\abbreviations{Abbreviations}{\hspace{0.1cm}The following abbreviations are used in this manuscript:\\

\noindent 
\begin{tabular}{@{}ll}
NTFP & non-thermal fixed point\\
BEC & Bose--Einstein condensate\\
IR & infrared\\
UV & ultraviolet
\end{tabular}}

\begin{adjustwidth}{-\extralength}{0cm}
\reftitle{References}


\end{adjustwidth}
\end{document}